# Band Gap Estimation of Multilayer 2D Semiconductor Channels Using Thin Graphite Contact

Sam Park[1], June Yeong Lim[1], Sanghyuck Yu[1], Kyunghee Choi[2], Jungcheol Kim[3], Hyeonsik Cheong[3], Seongil Im[1]

[1]vdWMRC, Department of Physics, Yonsei University, 50 Yonsei-ro, Seodaemun-gu, Seoul 03722, Korea.

[2]Electronics and Telecommunications Research Institute Reality Display Device Research Group, 218 Gajeong-ro, Yuseong-gu, Daejeon 34129, Korea.

[3]Department of Physics, Sogang University, 35 Baekbeom-ro, Mapo-gu, Seoul 04107, Korea.

Correspondence: Seongil Im (Semicon@yonsei.ac.kr)

## ABSTRACT

Band gap of monolayer and few layers in two dimensional (2D) semiconductors has usually been measured by optical probing such as photoluminescence (PL). However, if their exfoliated thickness is as large as a few nm (multilayer over ~5L), PL measurements become less effective and inaccurate because the optical transition of 2D semiconductor is changed from direct to indirect mode. Here, we introduce another way to estimate the bandgap of multilayer 2D van der Waals semiconductors; that is utilizing field effect transistor (FET) as a platform. We used graphene (thin graphite or multilayer graphene) contact for multilayer van der Waals channels in FET, because graphene contact would secure ambipolar behavior and enable Schottky contact barrier tuning of FETs with the assistance of top passivation. As a result, the bandgaps of multilayer transition metal dichalcogenides and black phosphorus in unknown thickness were successfully estimated through measuring the temperature-dependent transfer curve characteristics of prepared 2D FETs with graphene contact.

## INTRODUCTION

Two dimensional (2D) van der Waals semiconductors have been extensively studied for the last decade, regarded as one of the most important breakthrough materials toward future device technologies. [1-11] Many of transition metal dichalcogenides (TMDs) and black phosphorus (BP) would be the representatives of the 2D van der Waals semiconductors, which are mechanically exfoliated to be monolayer (1L), bilayer (2L), and even multilayer. [12–18] Band gap of monolayer and few layer in 2D semiconductors has been reported, measured by optical probing such as photoluminescence (PL). [18–21] However, if their exfoliated thickness is as large as a few nm (multilayer over 5L), PL measurements become less effective and inaccurate because the optical transition of 2D semiconductor is changed from direct to indirect mode. [22,23] At the moment, the energy band gap also becomes smaller with the semiconductor thickness. [23,24] Since the practical measurement of thick multilayer 2D semiconductor band gap is not easy, density function theory (DFT)-based calculations have been a main method to estimate the bandgap in general although somewhat special techniques using scan tunneling microscopy, absorption/reflectance spectroscopy using an optical microscope, and ion gel-gated FETs. [25–33] Here, we introduce another way to measure the bandgap of multilayer 2D semiconductors; that is also utilizing field effect transistor (FET) as a platform, which is but different from ion gel-gating. Our method is for transport gap which is more practical than optical one. We have fabricated multilayer 2D channel FETs with graphene (thin graphite) contact and top passivation for the present study, because graphene contact would secure ambipolar behavior and enable Schottky contact barrier tuning of FETs with the assistance of top passivation. [34–36] Bandgap could be estimated by achieving temperature-dependent transfer curve characteristics of prepared 2D FETs with graphene contact. (Here, we used to call “graphene” but in fact regard it as thin graphite). Although our method would effectively work for ambipolar 2D channel FETs, it is regarded that it is particularly useful for relatively low band gap materials such as black phosphorous (BP) and $MoTe_2$ [37,38], of which the band gap estimation appears difficult depending on their thickness.

## RESULTS AND DISCUSSION

Figure 1a~c shows the schematic band diagrams of FET channel with graphene contact under a negative drain/source voltage ($V_{DS}$). When gate voltage is negative, the diagram should be described as Figure 1a showing p-type hole conduction, while positive gate voltage changes the diagram to be like Figure 1c, n-type conduction. On the one hand, if the gate voltage is somewhere between two states to lead Fermi energy to the intrinsic level, intrinsic or highest Schottky barrier can be achieved as indicated in Figure 1b. Here, it is worth to note that the Fermi levels of graphene S/D and 2D TMD channel are always aligned at their van der Waals interface contact in thermal equilibrium, which means that the two Fermi energy ($E_F$) levels of graphite and channel in contact must be always the same at a certain temperature. [36] So, in our top contact FET, gate voltage will modulate the $E_F$ of 2D channel first, which then causes to modulate graphite's $E_F$; 2D channel and thin graphite are in contact. In principle, the highest barrier should be a half of real bandgap if it can be experimentally obtained. As a matter of fact, the value of Schottky barrier, $q\Phi_B$ between multilayer 2D channel and graphene is electric (E) field-dependent and also temperature-dependent. [39,40] Those two factor dependencies are simultaneously expressed by temperature-dependent transfer characteristics (drain current-gate voltage: $I_D$-$V_{GS}$), so that the bandgap of 2D channel may finally be extracted. In the present study, we initially took thin and thick $WSe_2$ channel layers as multilayer 2D semiconductor materials for the bandgap measurements in FET platform while thin $MoTe_2$ and BP channels were also probed later. As a result, we achieved very reasonable values from the measurements, and regard that our measurement method is both novel and practical as an important scientific tool to probe the bandgap estimation of multilayer indirect ambipolar 2D semiconductors, since PL or other

optical transition measurements might not easily carry out. We found this band gap estimation as only a piece of dual gated $MoTe_2$ device study, but the method seems to deserve consideration as a general principle of band gap measurement. [41]

An optical microscopy image in Figure 2a displays our first $WSe_2$ bottom gate FET ($WSe_2$ FET1) with graphene contact for source/drain (S/D) whose lead metal is Au. As shown in the inset schematic cross section of Figure 2c along with Figure 2a, thin hexagonal boron nitride (h-BN) layers were used for gate dielectric and top passivation, and gate (G) electrode was Pt on glass substrate. Thickness information of graphene and h-BN is provided in supplementary Figure S1, however, $WSe_2$ thickness is unknown and could not be clearly measured by atomic force microscope (AFM) or any other thickness profiler since it is encapsulated as a channel by other component materials (see the inset of Figure 2c). [Hence, we initially measured its thickness as shown in the inset AFM profile of Figure 2a before h-BN topping. The thickness appears ~12 nm.] Figure 2b and c show transfer (drain current-gate voltage: $I_D$-$V_{GS}$) and output (drain current-drain voltage: $I_D$-$V_{DS}$) characteristics of $WSe_2$ FET1, respectively. According to the transfer characteristics, the $WSe_2$ FET1 is an ambipolar device with a negative gate voltage of ~-2.5 V as its channel type transition point. The $V_{GS}$ points at ①, ②, and ③ are corresponding to the situations of Figure 1a, b, and c, respectively. Output curves also display ambipolar $I_D$ behavior of decrease and re-increase at $V_{DS}$= -1 V when $V_{GS}$ increases from -10 to +10 V. Gate leakage current, $I_G$ was very low below ~pA. Figure 2d shows temperature-dependent transfer curves of $WSe_2$ FET1 as obtained at several elevated temperatures for Schottky barrier extraction. [39,40,42] Using the following equation and transfer characteristics of Figure 2d, the plots of Figure 2e are constructed for each $V_{GS}$ condition.

$$I_{DS} = A^* T^2 \exp\left(-\frac{q\Phi_B}{kT}\right)\left[-1 + \exp\left(\frac{qV_{DS}}{kT}\right)\right] \quad (1)$$

, where $V_{DS}$ is -1 V we use, A* is Richardson's constant, k is Boltzmann constant, T is temperature (Kelvin) and we use $T^2$ instead of $T^1$ or $T^{3/2}$ because our flake is not a perfect atomistic 2D but ~10L-thick materials. [40,42,43] $q\Phi_B$ is Schottky barrier height. We can extract out the $V_{GS}$-dependent barrier heights which are the slopes of $\ln(I_{DS}/T^2)$ vs. 1/kT plots at the individual $V_{GS}$. Schottky barrier height is varied according to $V_{GS}$ as previously explained in Figure 1, and it is plotted in Figure 2f, where the peak value appears to be 0.58 eV at -2.2 V of $V_{GS}$. This means that the band gap of $WSe_2$ channel would be ~1.16 eV, twice of 0.58 eV which is the maximum Schottky barrier. At -2.2 V, the $WSe_2$ channel in our ambipolar $WSe_2$ FET1 must become an intrinsic semiconductor. Figure 3a~f present similar results to those of Figure 2a~f, as achieved from another $WSe_2$ FET ($WSe_2$ FET2) with different channel whose thickness value was again initially measured by AFM, to be 6~7 nm (inset of Figure 3a). $WSe_2$ FET2 shows minimum conduction point at a $V_{GS}$ of +3 V and its maximum $q\Phi_B$ near 3 V (~+ 2.2 V) unlike $WSe_2$ FET1 which has -2.5 V for its minimum conduction. The maximum Schottky barrier height appears to be ~0.72 eV, indicating that this $WSe_2$ channel has 1.44 eV as its band gap. So, it is expected that the channel thickness in $WSe_2$ FET1 would be larger than that of $WSe_2$ FET2. According to the literature, the estimated band gap values of thick and thinner $WSe_2$ channels in our FETs are regarded reasonable, to be 1.44 eV for 6~7 nm (9~10L) and 1.16 eV for 12 nm (17L), which is close to bulk. But we also attempted PL measurements directly probing on our two FET devices and other three $WSe_2$ flakes (with 1L, 7L, and 13L thickness on Si substrate and we took those for comparison). Although PL would not be effective for the band gap estimation of indirect gap $WSe_2$ layers which should have quite small exciton binding energy, the PL spectra of 7L $WSe_2$ flake and 10L (from our $WSe_2$ FET2) appear comparable each other at least (see Supporting Information Figure S2a and b for more details on the PL and Raman spectra of all our $WSe_2$ samples). [19,23,44,45]

Since we regarded our bandgap estimation results reasonable, we probed other 2D materials for their bandgaps: a few nm-thin $MoTe_2$ and BP. Figure 4a shows OM image of $MoTe_2$ channel FET with graphene and h-BN (also see the inset of Figure 4d for device cross section). Transfer and output curve characteristics of Supporting Information Figure S3 a and b display ambipolar behavior of $MoTe_2$ FET. Through the temperature-dependent transfer curve measurements of Figure 4b and its $\ln(I_{DS}/T^2)$ vs. 1/kT plots of Figure 4c, Schottky barrier was plotted as a function of $V_{GS}$ in Figure 4d, where the maximum height appears to be ~0.46 eV. The result indicates that the bandgap of $MoTe_2$ channel would be 0.92 eV although its exact thickness was unknown yet. We did not measure the thickness by AFM for this case, but instead attempted Raman probing on the $MoTe_2$ FET for more exact layer thickness. According to the low frequency Raman spectra of Figure 4e, we could precisely estimate the layer number or thickness of our $MoTe_2$ channel in FET, to be 9 layer (~6 nm) although typical high frequency spectra would not be much help (inset). Compared with the low frequency spectra from 2, 4, and 7 layer (L) flakes (Figure 4e), that of our 9L-thin $MoTe_2$ channel appears clearly distinguishable in respect of shear and breathing modes. [46] In fact, our estimated band gap seems like quite a reasonable value (0.92 eV) for 9L $MoTe_2$, according to literature. [18,22] In order to support our band gap results in 9L $MoTe_2$, we conducted the same experiment with another $MoTe_2$ layer which is as thin as ~4L (confirmed by low frequency Raman spectroscopy), since the PL results of such thin $MoTe_2$ (1L~4L) were reported to provide approximate band gap information. [18,22] According to the transfer curve characteristics in Figure 4f, the 4L-thin $MoTe_2$ FET shows quite strong p-type and weak n-type conduction. Resultant plot for maximum Schottky barrier in Figure 4g presents 0.55 eV, suggesting the transport band gap of 4L $MoTe_2$ to be ~1.1 eV. This value appears slightly higher than expected and reported PL results. [18,22] Using the 4L-thin ambipolar $MoTe_2$ FET, we again extracted the Schottky barriers for electrons (n-type channel in $MoTe_2$ FET) at $V_{DS}$ = 1 V (see the band diagrams in Supporting Information, Figure S4). According to Supporting Information, Figure S5, we can see that the maximum Schottky barrier heights are almost the same for both n- and p-channels (0.53 and 0.55 eV, respectively). These results again confirm that our band gap estimation method is reliable.

Thin BP channel was also taken for our bandgap measurements as shown in Figure 5a for its device image. In this case we could measure the channel thickness by AFM scan even after top passivation because conformal atomic layer deposition (ALD) $Al_2O_3$ was known to be an effective oxidation protector for BP, and ambipolar behavior-propeller as well. [47] The effective thickness of BP channel was thus measured, to be ~6 nm in Figure 5b where Raman spectra of our BP channel is also shown as inset. The maximum Schottky barrier height is extracted to be 0.21 eV from the plots of Figure 5c and d. (All of transfer and output curve information of BP FET are shown in Supporting Information Figure S6.) So, the bandgap of the present BP channel is determined to be 0.42 eV, which is quite comparable to the value in literature. [11] Our results on the BP band gap would not be changed by any doping in the material, however, if any type of surface oxidation was developed and changed/decreased the effective thickness of BP channel, our results might show a little higher band gap than expected. [37] Table 1 summarizes the bandgap values of the three 2D materials, which are dependent on thickness as cited from literature. According to the table, our measurements seem reasonable in respects of all the values of 2D channels we used for FETs.

As our last analysis to reconfirm our results, we also attempted output curve analysis [50] to estimate maximum $q\Phi_B$ and band gap ($E_g$) based on Eq. (1), where $V_{DS}$ is now forward bias, T is fixed as 300 K, and A* is defined by effective mass of charge. As the analysis details are found in Figure S7, $q\Phi_B$ results surprisingly display that peak $q\Phi_B$ exists for

graphite/$WSe_2$ and graphite/$MoTe_2$ Schottky junctions. The values appear to be 0.62 and 0.51 eV, which are quite consistent with results of Figure 2f and 4d.

In summary, we conducted the bandgap measurements of multilayer van der Waals semiconductor channels: $WSe_2$, $MoTe_2$, and BP in FET. Since those multilayer channels show ambipolar behavior with graphene contact which is able to tune Schottky barrier, their maximum Schottky barrier height is estimated through temperature-dependent transfer curve characteristics. The maximum Schottky barrier between multilayer channel and graphene is extracted near a $V_{GS}$ point where minimum $I_D$ is obtained. At the point, Fermi energy of each channel is at its intrinsic level, and the band gap is achieved as twice of Schottky barrier height. Our bandgap values from $WSe_2$ FET1, $WSe_2$ FET2, 4L-$MoTe_2$ FET, 9L-$MoTe_2$ FET, and BP FET were 1.16, 1.44, 1.1, 0.92, and 0.42 eV, respectively. The channel thickness or layer numbers were measured by AFM or low frequency Raman spectroscopy. The values from $WSe_2$ FETs were compared with those of PL measurements, and the results from two different measurements were found compatible. According to literature, our values are also regarded quite reasonable. We conclude that our bandgap measurement method is both novel and practical as an important scientific tool to probe the bandgap estimation of multilayer indirect 2D semiconductors.

**Table 1** Summary table for band gap information of mono-to-multilayer 2D semiconductors.

| Material | Theoretical $E_g$ Bulk~1L | Experimental $E_g$ | 1L Thickness |
|---|---|---|---|
| 9~10L-$WSe_2$[a)] | 1.2 eV ~ 1.64 eV [4,19,45] | 1.44 eV | 0.7 nm [19,45] |
| 17L-$WSe_2$[b)] | | 1.16 eV | |
| 4L-$MoTe_2$ | 0.81 eV ~ 1.2 eV [16,18,22] | 1.06 eV | 0.7 nm [9,16,48] |
| 9L-$MoTe_2$ | | 0.92 eV | |
| 6nm BP | 0.3 eV ~ 2.0 eV [11] | 0.42 eV | 0.53 nm [49] |

[a), b)] means $WSe_2$ in FET1, FET2. $E_g$ means band gap.

# METHODS

## Ambipolar FET device fabrication

Glass substrate was cleaned with acetone and ethanol using ultrasonicator. A 50 nm-thin Pt patterned gate electrode was deposited on the glass substrate through photolithography and DC sputter deposition. h-BN flake which was mechanically exfoliated by Polydimethylsiloxane (PDMS) was transferred on the Pt patterned gate as a gate insulator. Then exfoliated multilayer $WSe_2$ and $MoTe_2$ flakes were transferred on the h-BN as transistor channel, respectively. In the same way, exfoliated graphene as source/drain (S/D) contact was transferred on the channel flake and here S/D contact regions should have overlap with gate (G) region in consideration of gating effect on graphene. Finally, as a top passivation, another h-BN layer was transferred on top of device. In the case of Black Phosphorus (BP) channel FET, h-BN was replaced with 50 nm-thick $Al_2O_3$ which was deposited by Atomic Layer deposition (ALD). For measurement, Au lead pattern was deposited on S/D graphene by photolithography and DC sputter deposition.

## Device and materials characterization

Device characteristics were obtained in the dark by using a semiconductor parameter analyzer (4155C Agilent). For temperature dependent characteristics, a hot chuck was used in the probe station. Photoluminescence (PL) and Raman spectroscopy were taken with laser source of $\lambda$= 532 nm. The thickness of h-BN and graphene was measured by Atomic Force Microscope (AFM).

**ACKNOWLEDGEMENTS**

S.P. and J.Y.L. contributed equally to this work. The authors acknowledge the financial support from NRF (NRL program: Grant No. 2017R1A2A1A05001278, SRC program: Grant No. 2017R1A5A1014862, vdWMRC). J.Y.L acknowledges the tuition support from the Hyundai Motor Chung Mong-Koo Foundation.

**AUTHOR CONTRIBUTIONS**

S.P. fabricated $WSe_2$ and $MoTe_2$ devices and J.Y.L. fabricated BP devices; S.P. and J.Y.L. conducted data analysis; K.C. helped data analysis; S.Y. contributed with 3D scheme figure; J.K. and H.C. conducted PL and Raman spectroscopy; S.I. designed whole-device experiments.

## Figures

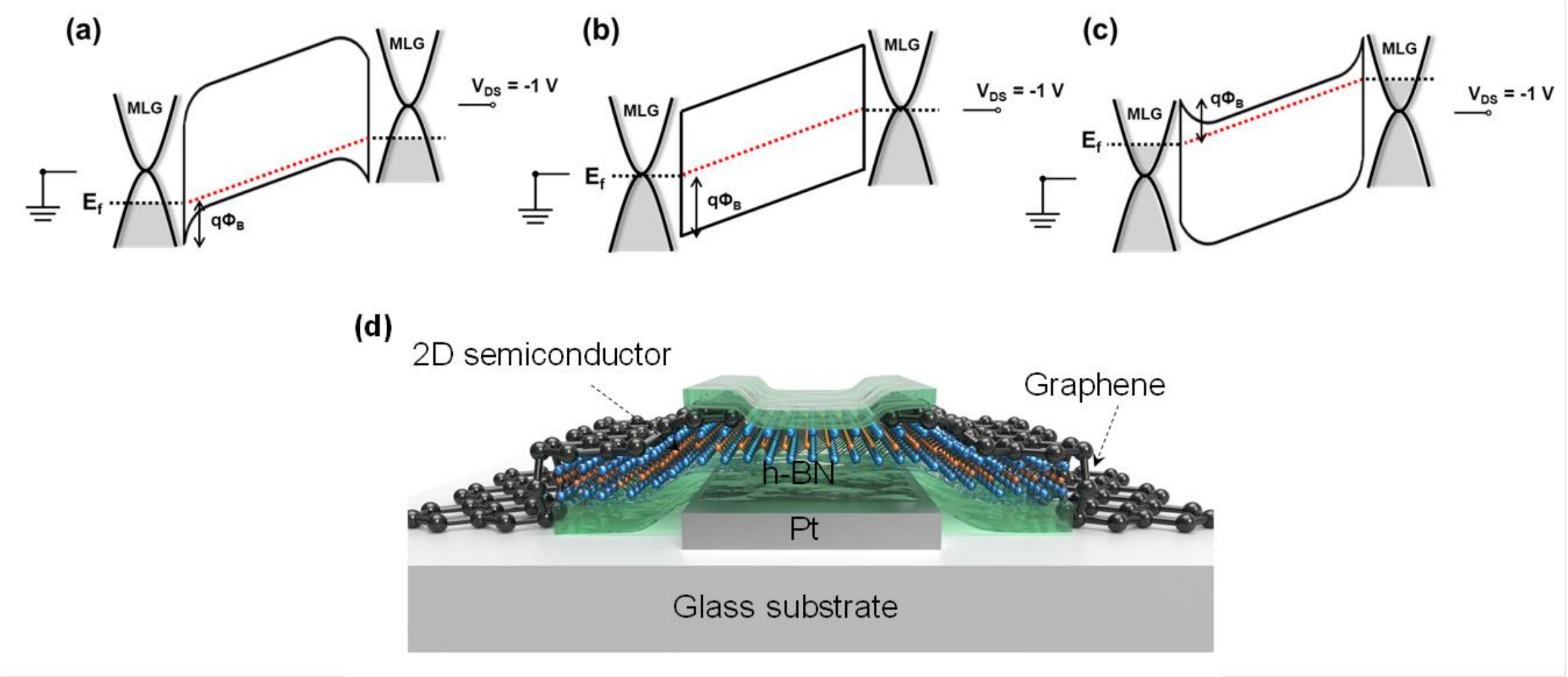

**Fig. 1** Schematic band diagrams of 2D semiconductor channel with multilayer graphene (MLG) or thin graphite contact under negative (-) $V_{DS}$. **a** figure corresponds to the situation under (-) $V_{GS}$ which makes p-channel. **b** figure is the case of neutral intrinsic semiconductor mode, where the Schottky barrier height $q\Phi_B$ becomes maximum. **c** figure shows the situation when positive (+) $V_{GS}$ is applied and thus n-channel forms. It should be noted that graphene/channel $q\Phi_B$ is tuned by $V_{GS}$ due to the semiconducting property of MLG. **d** 3D schematic view of 2D semiconductor device with graphene top contact. In the cases of **b**, we arbitrary describe the $E_F$ of MLG and there is no guarantee that the $E_F$ of MLG coincides to the Dirac point under a $V_{GS}$ which modulates 2D semiconductor to be intrinsic.

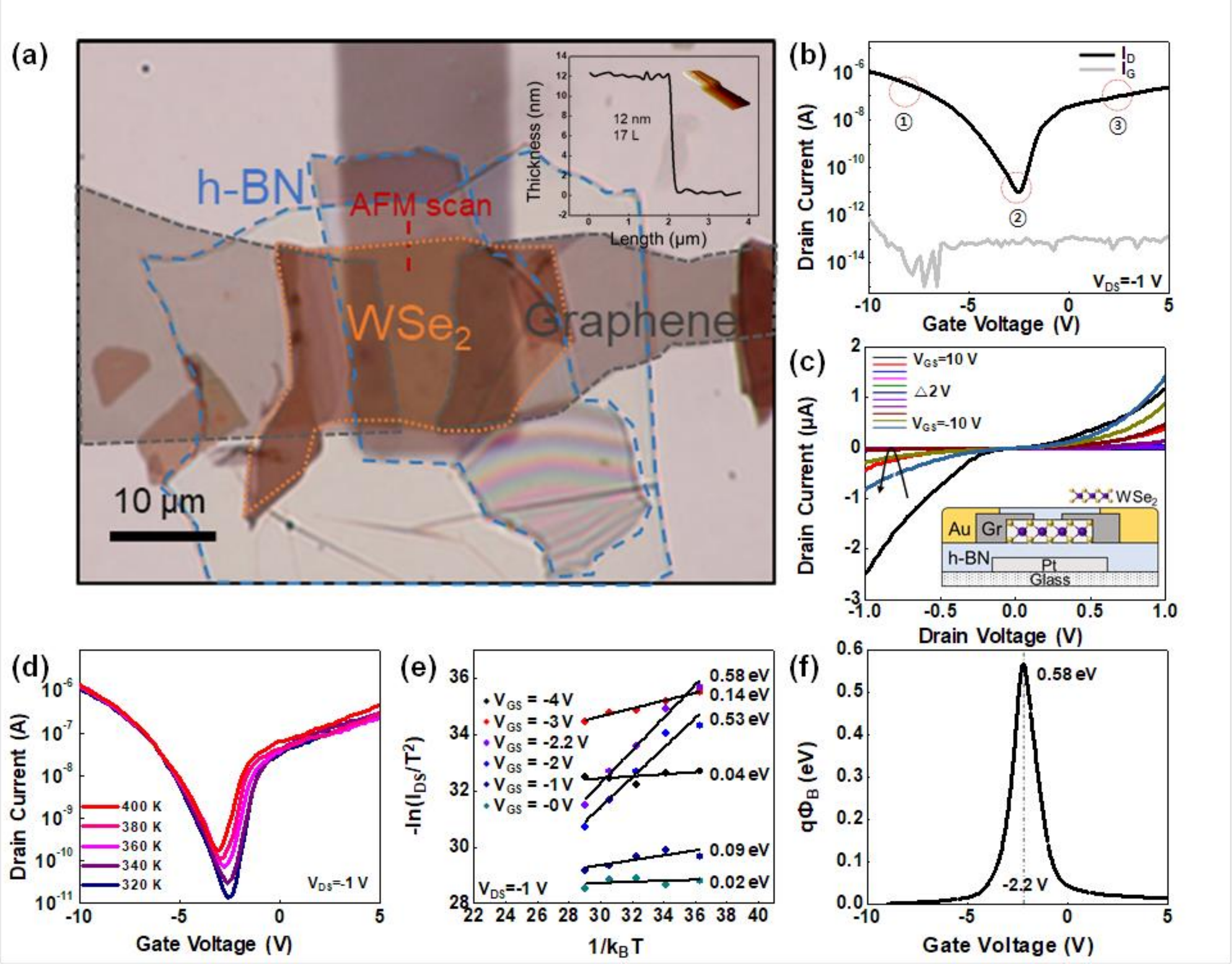

**Fig. 2** **a** Bottom illumination Optical Microscope (OM) image of $WSe_2$ FET1 with graphene contact and h-BN top passivation. Inset is AFM scan image for channel flake. **b** Transfer characteristic of ambipolar $WSe_2$ FET1 where $V_{GS}$ positions of ①, ②, and ③ correspond to **a**, **b**, and **c** band diagrams in Figure 1, respectively. **c** Output characteristics of ambipolar $WSe_2$ FET1. Note the inset device cross-section of $WSe_2$ FET1, where channel is embedded by h-BN dielectrics and graphene contact electrodes. **d** Temperature-dependent transfer curves as obtained at several elevated temperatures. **e** $\ln(I_{DS}/T^2)$ vs. 1/kT plots achieved from temperature-dependent transfer curves for example $V_{GS}$ locations. Slope of a plot represents the Schottky barrier height, $q\Phi_B$ at the $V_{GS}$. **f** $q\Phi_B$ plot as a function of $V_{GS}$. At -2.2 V, the maximum $q\Phi_B$ appears to be about 0.58 eV and resulting energy gap of channel $WSe_2$ becomes ~1.16 eV.

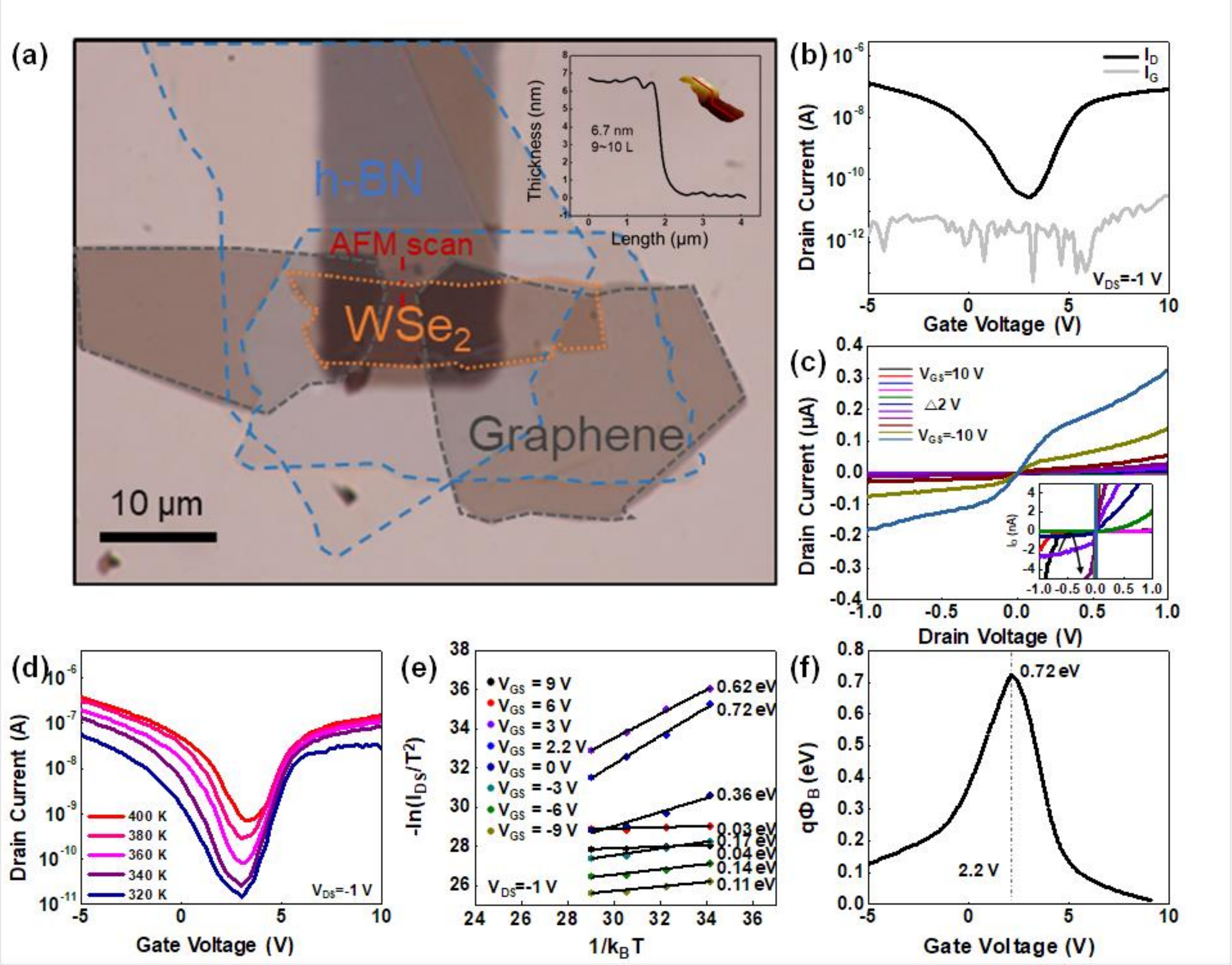

**Fig. 3** **a** Optical Microscope (OM) image of $WSe_2$ FET2 with graphene contact and h-BN top passivation. Inset is AFM scan image for channel flake. **b** Transfer characteristic of ambipolar $WSe_2$ FET2. **c** Output characteristics of ambipolar $WSe_2$ FET2. Note the zoomed inset curves show ambipolar output behavior of $WSe_2$ FET2. **d** Temperature-dependent transfer curves as obtained at several elevated temperatures. **e** $\ln(I_{DS}/T^2)$ vs. 1/kT plots achieved from temperature-dependent transfer curves for several $V_{GS}$ locations. **f** $q\Phi_B$ plot as a function of $V_{GS}$. At +2.2 V, the maximum $q\Phi_B$ appears to be about 0.72 eV and resulting energy gap of channel $WSe_2$ becomes ~1.44 eV.

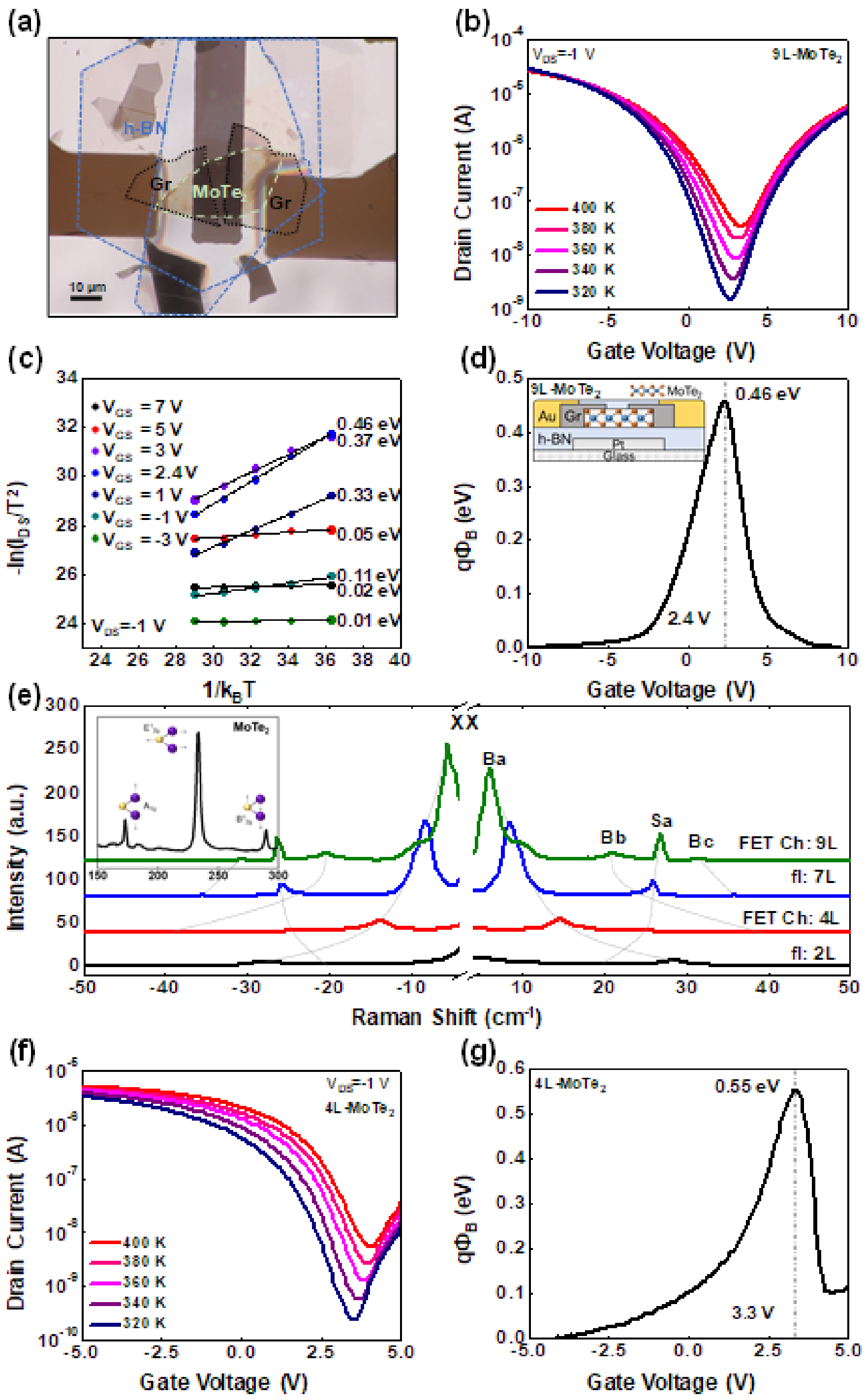

**Fig. 4** **a** Bottom-illumination OM image of multilayer 9L-$MoTe_2$ FET. **b** Temperature-dependent transfer curves of $MoTe_2$ FET as obtained at several elevated temperatures. **c** $\ln(I_{DS}/T^2)$ vs. 1/kT plots achieved from temperature-dependent transfer curves for several $V_{GS}$ locations. **d** $q\Phi_B$ plot as a function of $V_{GS}$. At +2.4 V, the maximum $q\Phi_B$ appears to be about 0.46 eV and resulting energy gap of channel 9L-$MoTe_2$ becomes ~0.92 eV. Inset is schematic

device cross section view of $MoTe_2$ FET. **e** Low frequency Raman spectra of $MoTe_2$ flakes (fl) and our $MoTe_2$ FET channel, recorded at $E_L$=2.33 eV in the parallel (XX) configuration. By comparing peak positions (Breathing and Shear modes: Ba, Bb, Bc, and Sa) of each flake, our channel turns out to be 9L-thick. Inset is high frequency Raman spectra of $MoTe_2$ channel, which show typical spectra of multilayer $MoTe_2$. **f** Temperature-dependent transfer curves of 4L-$MoTe_2$ FET as obtained at several elevated temperatures. **g** $q\Phi_B$ plot as a function of $V_{GS}$. At +3.3 V, the maximum $q\Phi_B$ appears to be about 0.55 eV and resulting energy gap of channel 4L-$MoTe_2$ becomes ~1.1 eV.

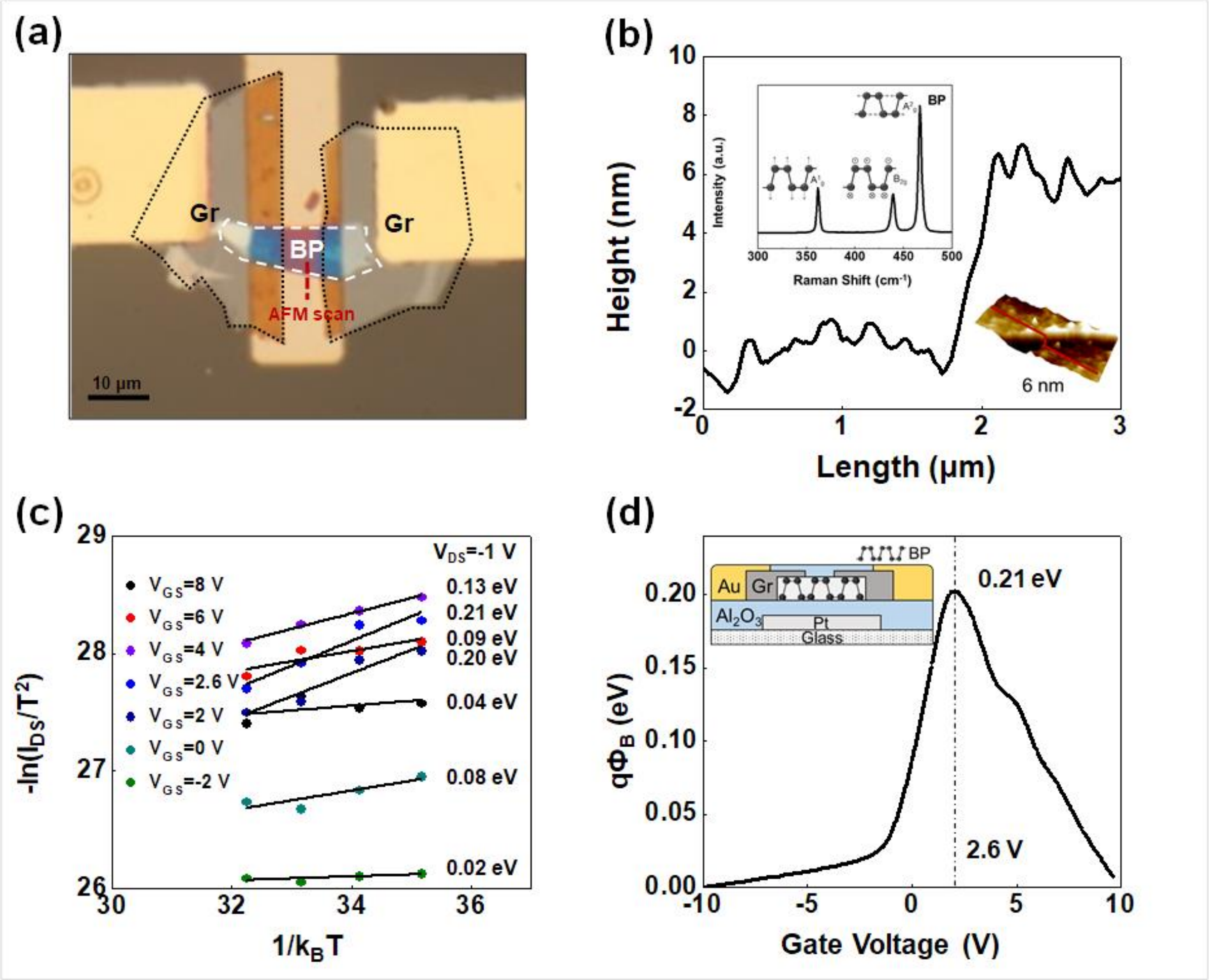

**Fig. 5** **a** Top-illumination OM image of multilayer black phosphorus (BP) FET. **b** AFM image of 6 nm thick BP channel flake. From inset Raman spectra, typical spectra of multilayer BP are observed. **c** $\ln(I_{DS}/T^2)$ vs. 1/kT plots achieved from temperature-dependent transfer curves for several $V_{GS}$ locations. The slope of each plot means Schottky barrier height. **d** $q\Phi_B$ plot as a function of $V_{GS}$. At +2.6 V, the maximum $q\Phi_B$ appears to be about 0.21 eV and resulting energy gap of channel BP becomes ~0.42 eV. Inset is schematic device cross section view of BP FET. Here, h-BN was replaced with Atomic Layer Deposited (ALD) $Al_2O_3$ layers for gate insulator and passivation.